# HYBRID TRUST MODEL FOR INTERNET ROUTING

Pekka Rantala, Seppo Virtanen and Jouni Isoaho

Turku Centre of Computer Science,

University of Turku, Finland

{Pekka.rantala, seppo.virtanen, jouni.isoaho}@utu.fi

http://www.it.utu.fi

## ABSTRACT

*The current Internet is based on a fundamental assumption of reliability and good intent among actors in the network. Unfortunately, unreliable and malicious behaviour is becoming a major obstacle for Internet communication. In order to improve the trustworthiness and reliability of the network infrastructure, we propose a novel trust model to be incorporated into BGP routing. In our approach, trust model is defined by combining voting and recommendation to direct trust estimation for neighbour routers located in different autonomous systems. We illustrate the impact of our approach with cases that demonstrate the indication of distrusted paths beyond the nearest neighbours and the detection of a distrusted neighbour advertising a trusted path. We simulated the impact of weighting voted and direct trust in a rectangular grid of 15\*15 nodes (autonomous systems) with a randomly connected topology.*

## 1. INTRODUCTION

The Internet is facing many severe problems, many of them arising from the age of the underlying core solutions of the network. Multimedia and entertainment applications like high definition video on demand services are placing higher demands for sustainable bandwidth and transfer rate. The drive towards offloading computation tasks to processing clouds is a considerable near-future bandwidth challenge for the Internet. Local and regional traffic surges incurred by malware are also a severe threat for data transportation stability.

All of these trends put great demands on the routing system of the Internet: the vast amounts of data moving between end users must be routed reliably, securely and efficiently across the world despite all the encountered challenges. It is clear that the entire Internet and the systems making it operational cannot be updated instantaneously for current and future requirements. Therefore, solutions that are able to function on top of existing systems and are at the same time capable of scaling up to future standards and requirements are needed.

The Internet is formed of distributed autonomous entities (Autonomous Systems, ASes) that communicate with each other by exchanging messages with Border Gateway Protocol (BGP). As such, the security of the inter-AS routing system is an essential factor in ensuring reliable data transportation in the network. Because all routers co-operate in the collection and distribution of topology information, they can easily propagate spoofed routing information to each other. Illegitimate routing information can be propagated both by unintentional routing misconfigurations [1] and attacks by a malicious adversary [2]. Both types often cause traffic instability and reachability problems for at least a small subset of ASes on the Internet.

Trust is a complex concept that is difficult to stringently define. A wide variety of definitions of trust have been put forward, many of which are dependent on the context in which interaction occurs, or on the observer's subjective point of view [3]. Trust in inter-domain routing is traditionally based on the personal relationships between human network operators who manage ASes. Especially among backbone and other large transit ASes, many ISPs have already developed trust relationships through personal and repeated contact at meetings (network operator's groups in different regions: e.g. NANOG [4] and AfNOG [5]). Additionally, trust between ASes can be formulated through successful business relationships or known adherence to best common security practices [6].

In this study we define the *trust scope as to route traffic via most desired path, defined by weighting components, and recommender trust as the trust to advertise trusted paths honestly.* We propose a novel trust model for inter-AS BGP routers. The trust model is developed considering future Internet challenges in a multi-domain environment [7]. In our approach, the trust model is defined by combining voting and recommendation to direct trust estimation for neighbour routers located in different autonomous systems. Our approach combines trust philosophy from three different trust model classes, for which reason we call it the *hybrid* trust model or hybrid T-BGP. This article continues with an overview of most important related work. Our trust management model and illustrative cases and simulations on the model's impact are presented from section 3 onwards. In section 5 we draw final conclusions and summarize the work presented in this study.

## 2. RELATED WORK

In order to fight against fallacious inter-AS routing messages, the research community has focused on two broad classes of possible solutions. On one hand, researchers have proposed modifications to the actual BGP protocol and the addition of a centralized public key infrastructure (PKI) or a routing registry [8–10]. Adoptions of these schemes have so far failed for a variety of reasons: the potential for out-of-date central databases, the inability for incremental deployment or the prohibitive overhead of expensive cryptographic operations. On the other hand, researchers have suggested many analytic tools to mitigate the effects of erroneous updates locally at each AS [11–13]. While many have proven useful, many detect only specific types of misconfigurations and attacks. These locally-deployed tools also do not take advantage of AS coordination to resolve inter-domain routing problems from multiple vantage points. None of these studies have utilized notion of trust in their models.

Yu et al. [15] approach the issue of misguided trust between ASes by deploying a novel distributed reputation protocol that relies on existing trust relationships between network operators. The approach mimics real-world trust relationships in social networks, putting more emphasis on information gathered from more trusted peers. The approach does not directly require alterations to the BGP protocol but the trust is built as inter-AS co-operation.

Ning Hu et al. [16] describe an approach in which reputation is calculated with the Bayesian probability model using an alliance of domains (ASes). In comparison to a fully distributed reputation model, their model exhibits lower storage and communication overhead.

### 2.1. Trusted BGP Routing

Liwen He [18] has presented a trust model for inter-domain routing that is reviewed in this subsection. He proposes a tree structure (dotted lines in the extended version in Fig. 2) for evaluating *direct trust*. Direct trust is divided into two major trunks, that is, two main trust attributes*: Inherent trust* and *Observed trust*. These trust values are all determined locally based on analytical observations tools on border routers, or data provided directly to the system by the local administrator. To incorporate his model into the BGP protocol, He proposes an enhancement to it called T-BGP. In T-BGP, all routers send their inherent trust information including political, financial, technical, historical, and operational information to their peers (leafs in the Fig. 2), and evaluate the initial values for trust rates for all their neighbours during the initialization phase. This requires an open and transparent international standard which can help transfer the above information into a single representative value of inherent *trust rate.*

*Trust rate* is a measure that quantifies the dynamics of trust relationships between routers. It is a numerical value that represents trustworthiness between two network entities subject to a specific property that is represented by a decimal number between 0.0 and 1.0. 0 means complete distrust and 1 means complete trust. 0.5 means uncertainty (default value). Entity A can calculate its *universal trust rate* for B using the following equation:

$$T_{A \to B} = \omega_1 * I_t + \omega_2 * O_t \in \{0, 1\}, \qquad (1)$$

where $I_t$ is A's inherent trust rate for B and $O_t$ is A's observed trust rate for B. $\omega$s are weighting coefficients. $I_t$ and $O_t$ are also weighted sums of their components. $I_t, O_t \in \{0, 1\}$ as well as $\omega_1, \omega_2 \in \{0, 1\}$. The trust relationship between two network entities is asymmetric.

Peering routers will periodically exchange trust-related information to maintain the trust relationship with each other. In a realistic routing operation, when the utilisation rate of a router is too high/low or the dropping/delivery rate is too high/low, the router might be under attack or congested. Therefore, the neighbouring router may want to degrade its trust rate components "Router Utilization" and "Packet Dropping" in Fig. 2 because this router is no longer trusted for reliably delivering traffic.

In He's model [18], a router uses its trust rate metric to set up routes that are physically secure and that conform to a certain trusted set of IP routing operational policies. We express this as *Normalised routing criterion with trust rate* (Cost):

$$C_t = C/T. \qquad (2)$$

$C$ in expression 2 is the original routing criterion value for router A to make a routing decision. The lower the value for $C$, the more preferable the corresponding routing decision is. $T$ is the trust rate value (between 0 and 1) in router A for its neighbouring router B. $C_t$ denotes the normalised routing criterion value with the trust rate used by the router to make the routing decision.

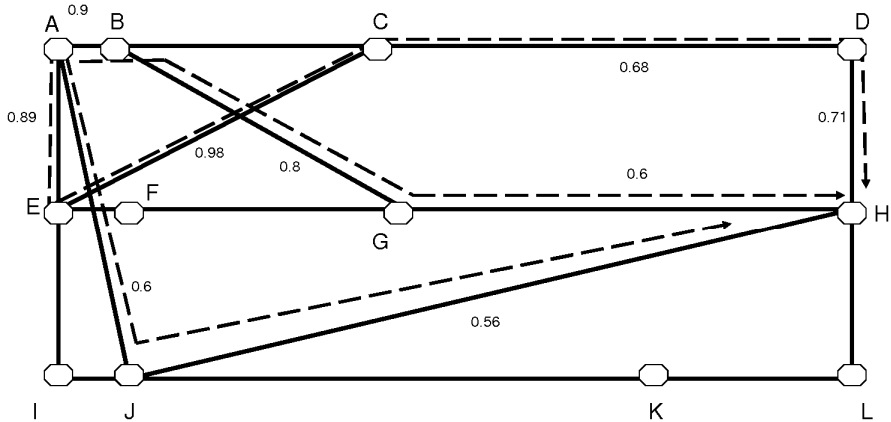

Figure 1. Routed network topology with link trust rates

The following example of computing the costs of the paths between nodes A and H is used to illustrate the use of the T-BGP protocol. The network topology used in this example is shown in Fig. 1 and is originally from Tanenbaum's book [19].

In Fig. 1, there are three alternative paths marked with dotted lines from node A to H. The links on the paths have trust rates (defined only for demonstrative purpose) observed from A's direction. A has three neighbouring routers: B, E and J. At the initialisation stage, routers B, G and H are used as an example. G knows it is one hop away from its neighbour H, and knows its trust rate value for H, that is, $T_{G \to H} = 0.60$. Thus, the normalised cost from G to H is: $C_{T(G-H)} = C/(T_{G-H}) = 1/0.6 = 1.67$. Next, G will send B (and F) the routing path information for reaching H, G-H, and the normalised cost, 1.67. B can then calculate its normalised cost to H. B knows it is one hop away from G, and it knows its trust rate value for G, ($T_{B-G} = 0.80$).

All three aggregate costs of the paths are calculated similarly with equation:

$$C_{path} = \sum_{l \in P} C_l = \sum_{l \in P} \frac{1}{T_l} = \frac{1}{T_1} + \frac{1}{T_2} + \ldots + \frac{1}{T_N}, \qquad (3)$$

where $C_l$ and $T_l$ are cost and trust rates of a single link on the whole path. The results are tabulated in table 1.

Table 1. Path cost values with direct trust

| Path | Path Cost |
|---|---|
| $A \to J \to H$ | $\frac{1}{0.6} + \frac{1}{0.56} = \mathbf{3.5}$ |
| $A \to B \to G \to H$ | $\frac{1}{0.9} + \frac{1}{0.8} + \frac{1}{0.6} = 4.0$ |
| $A \to E \to C \to D \to H$ | $\frac{1}{0.89} + \frac{1}{0.98} + \frac{1}{0.68} + \frac{1}{0.71} = 5.0$ |

Based on the calculation, A may decide to choose the path A-J-H instead of the path A-B-G-H to reach H. In this case, the router rationally selects the path with the highest overall trust, according to its trust-aware routing policy. In this case, the most trustworthy path is also the shortest path (in number of nodes).

The challenge in Liwen He's trusted routing [18] is simply the trust. In the distributed T-BGP concept discussed in section 2.1, node A gets the trust values of the paths from its neighbours. Hence, the costs calculated with expression 3 are correct *only if* the intermediate nodes on the path transmit the trust rates *reliably* to their neighbours. In other words, we cannot fully trust the trust rates that have traversed via distrusted nodes. Therefore this trusted routing mechanism would be more feasible under centralised control which, on the other hand, is not efficient and feasible in the distributed Internet.

## 3. HYBRID TRUST MODEL

We use He's trust model as the starting point for defining our *hybrid trust model* and combine to it the advantages of *recommendation* based trust introduced by Abdul-Rahman and Hailes [20]. We also partially adopt a trust model considering voting by Yu et al. [15] and Hu et al. [16]. With this approach we obtain a complete and useful solution for trusted routing in a distributed environment but that can still be maintained and operated locally.

The *hybrid* trust model requires the addition of a new measure, *voted trust*, to the trusted routing model suggested by He. We also add a recommender *coefficient* to the trusted path equation given in Eq. 3. The components of the novel trust model are presented in Fig. 2. In this figure, the direct trust tree proposed by Liwen He [18] is shown using dotted lines. The majority of our Hybrid trust model consists of the same components as He's model. We review the voting component in section 3.1.

In Fig. 2 there are also two new branches for inherent trust*: Ecological* and *AS relation*. The Ecological component bases on the ever increasing ecological perspective of commercial decisions and can be considered as a notable part of politics and economics in future. AS relation includes AS tier, size and commercial relation to the target AS. These are introduced by Hu et al. [16].

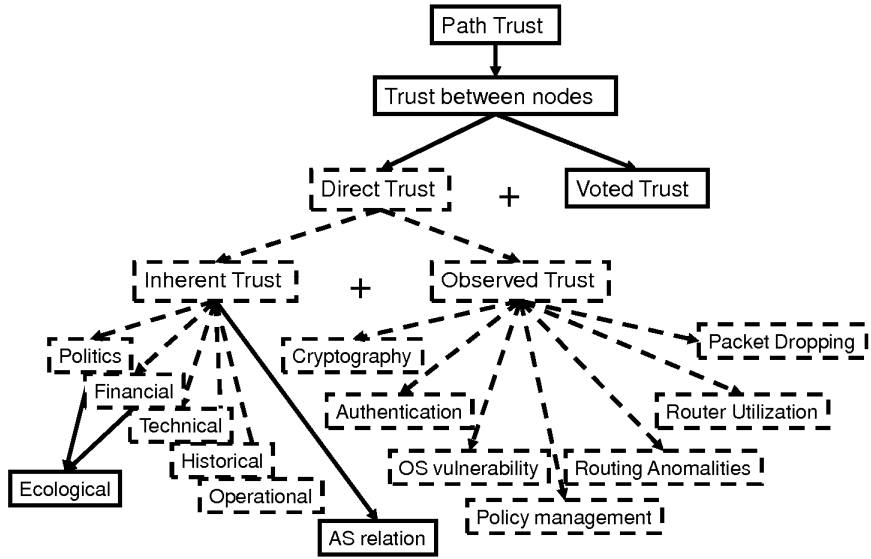

Figure 2. Components of trust in Internet routing

By adding the voting term to Eq. 1 we obtain Eq. 4 which sums up the components of the trust. In this equation, $I_t$ is A's inherent trust rate for A to trust B, and $O_t$ is the observed trust rate, respectively. $V_t$ is the weighted average of voted trust rates *from B's neighbours*. $\omega_1$, $\omega_2$ and $\omega_3$ are weighting coefficients the tuning of which is beyond the scope of this research. They satisfy the following conditions: $\omega_1$, $\omega_2$ and $\omega_3 \in \{0, 1\}$ and $\omega_1 + \omega_2 + \omega_3 = 1$.

$$T_{A \to B} = \omega_1 * I_t + \omega_2 * O_t + \omega_3 * V_t \quad (4)$$

### 3.1. Voted Trust

The concept of voting on the reputation of a neighbour is adopted from P2P social networks [16, 17]. Traditionally, the trust in the inter-domain routing context bases on the business contracts between domains. This component is included in the inherent and observed trust in Fig. 2.

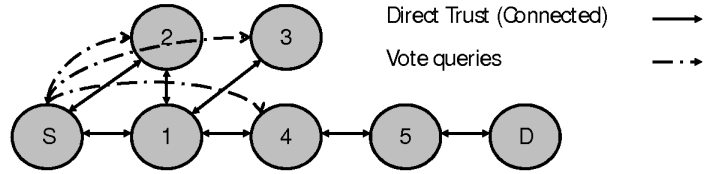

Figure 3 Vote querying

Voting among neighbours increases the reliability of the result, because every entity in the network is a separate AS. A shilling attack becomes more tolerable when the majority of voters is honest. The voting protocol can be implemented as an overlay network above the T-BGP-protocol.

In our approach we restrict the area of observation to the second degree neighbours. We call this area the neighbourhood. That means querying all the neighbours of a neighbour about the reputation of a neighbour (for example, ASes 2, 3 and 4 for AS 1 in Fig. 3). This kind of limitation of the trust information paths supports scalability and reliability of the system. Mathematically, the voting is a weighted average of the votes of the second degree neighbours. The $T_i$ weight factors in Eq. 5 are trust rates for the corresponding AS. If the AS is not a first degree neighbour we may define some relatively low constant weight factors for the second degree neighbour. This way we can put more weight on the votes of the first degree neighbours and more trusted voters. $V_i$ is the individual vote of an AS and is between 0.0 and 1.0.

$$V_t = \sum_{i=1}^{n}(V_i * \frac{T_i}{\sum_{i=1}^{n} T_i}) \in \{0,1\} \tag{5}$$

## 3.2. Recommendation Coefficient

In an ideal centralised system, the recommender coefficients would become factors to every term on a trusted path $(C_{path} = 1/T1 + 1/(T1*T2) + …)$. However, it is useless for a single node to maintain a large database for the link trust rates of the whole Internet because the information traverses through potentially distrusted nodes. Also due to scalability, the most important trust rates, the links to the neighbours, are the only values a single node uses for evaluating the path trusts. This makes the calculation this simple:

$$C_{path} = \frac{1}{T_1} + \frac{1}{T_1} C_{1.hoppath} = (1 + C_{1.hoppath}) * \frac{1}{T_1} \tag{6}$$

$C_{path}$ is now the overall trust rate for the whole path. $T_1$ is direct trust of the neighbour node (first link). The $T_1$ is used again as the recommender coefficient of the neighbour node (second $T_1$ in Eq. 6). A slightly different weighting of the recommender coefficient could be used, but because of the simplification we use the same term in this protocol twice. $C_{1.hoppath}$ is the overall trust of the rest of the path transferred and aggregated by the neighbour node.

### 3.3 Protocol Characteristic

The novel distributed hybrid T-BGP protocol described in this section utilizes several advantages from different trust schemes.

1. *Difficult to shill the entire system:* The distributed voting and weight factor α forces a node's vote to have smaller impact for distrusted nodes. The effect of a single malicious vote, even coming from a highly regarded AS, will have only a small impact on direct peers and a tiny effect on ASes two hops away in the overlay. α can be tuned to weigh local analysis tools i.e. direct trust to the neighbour when it seems more reliable than voting.
2. *Incrementally deployable:* Our proposed solution makes no changes to the de facto BGP protocol or the packet format of current BGP path announcements. The voting and reputation system is completely disjoint from BGP's control plane and can be adopted at any time by network operators. The system would also function regardless of the choice of local detection tools run at each AS.
3. *Confidentiality of AS relationships:* ASes closely guard confidential information about their business relationships with other ASes which can be compromised by assessing trusts of the neighbours. However, direct trusts and voted trusts are aggregated and averaged before being propagated further. In addition, cryptographic mechanisms can be utilised to encrypt and sign the votes cast in the overlay and provide security such that the votes will be unverifiable to a third party.
4. *Distributed local operation:* The lack of scalability hinders deploying most of the centralised trust management approaches studied earlier. Hybrid T-BGP operates only on neighbourhood of ASes and aggregates the trust rates for the whole paths. Deploying this protocol does not increase the processing overhead in the nodes when the network grows.

In general, our reputation based model is not inherently capable of detecting BGP misconfigurations and attacks itself. It is only useful when deployed in conjunction with the collection of available tools for debugging local networks. The positive trade-off, though, is that the architecture itself is agnostic to a specific problem in question, allowing any trust-related proposition to be weighted and voted on. With further research and tuning the trust components smartly, the protocol works well if a minority of participants act maliciously.

## 4. HYBRID TRUST MODEL SIMULATION CASES

In this section, we present three illustrative cases. First, we give a simple example of how the routing decision changes compared with He's T-BGP. In the second case, we assume that the second $T_1$ in Eq. 6 is derived from different factors than the first one and show the impact of this recommendation coefficient. Then we study the impact of making variations to the weighting of terms in Eq. 4. Specifically, we show how weighting the trust value obtained from voting in the neighbourhood improves trust detection as a function of neighbour count in comparison to using just direct trust.

### 4.1 Routing Decision

As our first example, we calculate new trust rates for the paths in Fig. 1. Node A gets the trust rates from neighbouring nodes B, E and J for their path to H similarly as in our previous example as shown in table 1. In our trust model, the impact of recommendation decreases the trust of the paths recommended by distrusted neighbours. In this case, the trust of path via node J decreases below the path of node B. When the path costs are recalculated with Eq. 6, the most desirable path changes to A-B-G-H (table 2). Hence, with the recommendation coefficient the indication of distrusted paths is improved and it extends beyond the nearest neighbours.

Table 2. Incrementing path cost values with recommender trust

| Path | Path cost |
|---|---|
| $A \to J \to H$ | $(1 + \frac{1}{0.56}) * \frac{1}{0.6} = 4.64$ |
| $A \to B \to G \to H$ | $(1 + (\frac{1}{0.8} + \frac{1}{0.6})) * \frac{1}{0.9} = \mathbf{4.35}$ |
| $A \to E \to C \to D \to H$ | $(1 + (\frac{1}{0.98} + \frac{1}{0.68} + \frac{1}{0.71})) * \frac{1}{0.89} = 5.51$ |

**4.2 Changing Trust Rates**

In this case we compare our model and Liwen He's model in terms of path trust behaviour when trust rates are changing. The difference between the models is exposed when a distrusted neighbour is advertising a trusted path. The slopes in Fig. 4 (Trust variation) represent changes in trust of a neighbour and the path behind it.

The solid line in Fig. 4 (Trust variation) presents linearly decreasing trust from 1.0 to 0.1 between source S and the neighbour node. The dotted line in this figure represents the total *expected* trust of the path behind the nearest neighbour (mediated by the neighbour). The chosen values of this exponentially growing set (0.05...0.33) are based on typical values of the trust of a single node (=0.5) and the typical path length in Internet (=5).

The total calculated path costs for the previous trust rates are presented in Fig. 4 (Total route cost variations). Calculations were made using Equations 3 and 6. In the figure it is seen that, with the recommendation coefficient, the total cost starts rising immediately as the trust of the neighbour decreases. By contrast, the total cost without the recommendation coefficient starts to decrease following the trust of the rest of the path. The trust rates of the path behind the nearest neighbour are transferred via the neighbour whose trust is decreasing. With the recommendation we are able to control the mediated values in this kind of scenario.

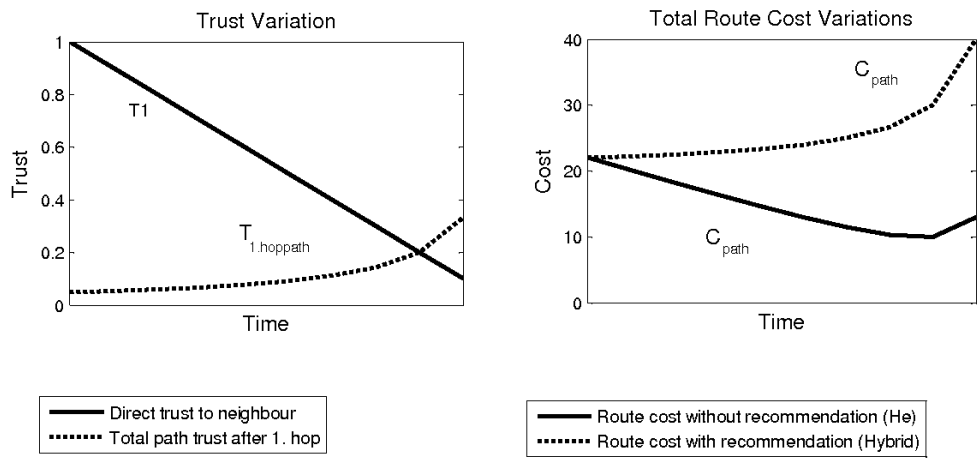

Figure 4. Behaviour of trust

International Journal of Computer Networks & Communications (IJCNC) Vol.3, No.3, May 2011

### 4.3 Weighting Voted and Direct Trust

Strong presumptions about interpretation and behaviour of the direct trust are required in this simulation. We use a five step scale for translating the trust. The steps are strong *distrust*: 0..0.2, *weak distrust*: 0.2..0.4, *neutral*: 0.4..0.6, *weak trust*: 0.6..0.8 and *strong trust*: 0.8..1.0. In this simulation we presume that the direct trust rates given to the neighbours are distributed normally around the average values 0.7 for the trusted nodes and 0.3 for the distrusted ones. The variation of the evaluation results is presumed to be 0.2. We define the failed detection of trust/distrust when a trusted/distrusted node gets a neutral trust rate i.e. over 0.4/under 0.6. These exact numerical presumptions are only examples in order to illustrate the behaviour of the protocol. We simplify Eq. 4 to the form *αT$_d$ +(1−α)V$_t$*, where $T_d = \omega_1 * I_t + \omega_2 * O_t$. Now, the weighting factor α divides the resulting trust between voted and direct trust component similarly as in the trust tree in Fig. 2.

In Fig. 5, the behaviour of trust detection failure is plotted as a function of voted trust weighting and neighbour count. The failure of detection is simulated by making 20 % of the nodes (AS's) in a rectangular 15*15 grid distrusted, and distributing the distrusted nodes randomly and uniformly in the grid. The links between nodes are connected vertically, horizontally and diagonally so that the maximum number of possible first degree neighbours is eight. Then, we removed links randomly in order to inspect the behaviour with fewer neighbours. The topology does not resemble the Internet very well, but the form of the resulting plane is not dependent on the exact topology characteristics. In the simulation, the trusted nodes query votes from the neighbours of neighbours, and the trusted nodes provide their derived trust values to the neighbour in response. If they do not yet have the voted result, they send the direct trust value to the neighbour. The direct trusts were randomly given from the normal distributions. We also presume that the vote from a *distrusted* node is randomly generated from uniform distribution on the unit interval. The failure of the detection of the trust is defined by setting the trust rate lower than 0.6 for a trusted node and correspondingly over 0.4 for a distrusted node.

From Fig. 5 we can first note that the mixing of voting and direct trust evaluation is simply beneficial. In the simulations, the lowest ratio of failure detection is always observed when the weighting between voted trust and direct trust is between 40% and 60%. Naturally, when we have more neighbours, 6.5 on average, it is wise to prefer voting (from many neighbours of neighbours). In this case the optimal value of alpha is around 0.4. Vice versa, when the amount of neighbours is around 2.2 and they are possibly distrusted voters, the optimal weighting factor for the own direct evaluation is 0.6. The resulting plane is rotated and rises a little bit when the neighbourhood is shrinking. Simply, voting is more efficient with more voters.

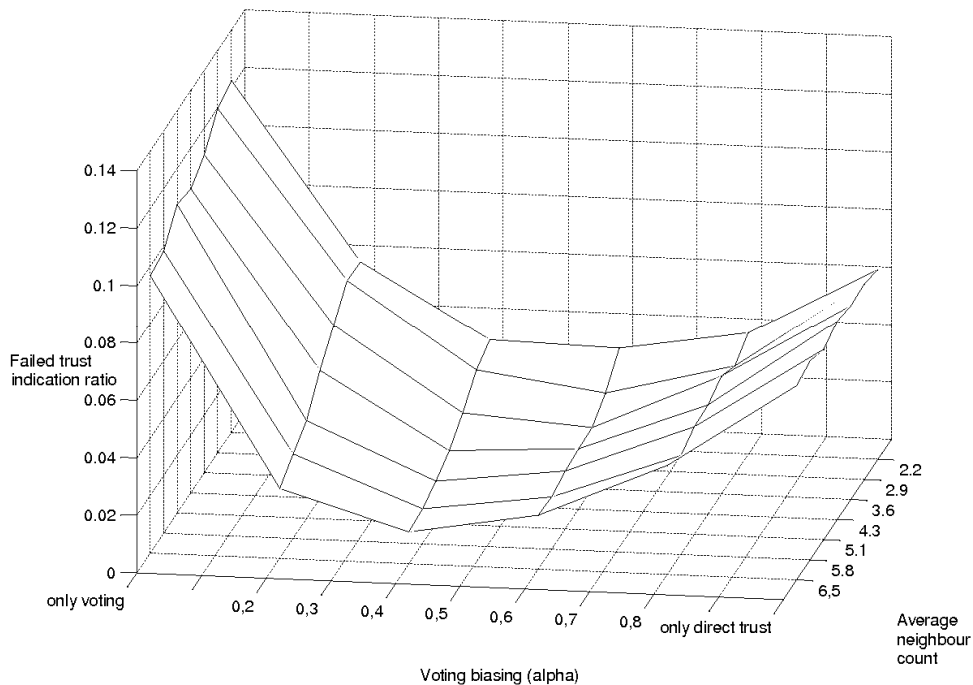
Figure 5. Behaviour of voted trust weighting

## 5. CONCLUSIONS

To improve the trustworthiness and reliability of the network infrastructure, we proposed the Hybrid trust model to BGP routing. In our approach, voted trust values are received from trusted neighbour nodes located in different Autonomous Systems. The received votes are combined with local trust information, and from the combination a final trust metric for each routing decision is derived. The voting and recommendation property enhances the reliability in distrusted environments where a neighbour router may send fraudulent trust rates. We illustrated the impact of our approach with cases that demonstrate the indication of distrusted paths beyond the nearest neighbours and the detection of a distrusted neighbour advertising a trusted path. We simulated the impact of weighting voted and direct trust in a rectangular grid of 15*15 nodes (Autonomous Systems) with a randomly connected topology. In our simulations, the lowest ratio of failure detection was always observed when the weighting between voted trust and direct trust was between 40% and 60%.

**ACKNOWLEDGEMENTS**
This work was supported by TEKES as part of the Future Internet programme of TIVIT (Finnish Strategic Centre for Science, Technology and Innovation in the field of ICT).

**Authors**

**Pekka Rantala** is a research scientist working towards a Ph.D. at the University of Turku, Finland. His research interest includes multi-agent systems and routing and trust management issues on both Network-on-Chip and Internet. He has a M.Sc. degree from the University of Turku.

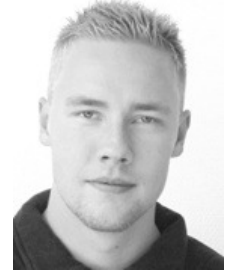

**Seppo Virtanen** received his BSc in applied physics, MSc in electronics and information technology (1998), and DSc (Tech.) in communication systems (2004) from the university of Turku (Finland). Since 2009, he has been adjunct professor of embedded communication systems in University of Turku. He is Editor-in-Chief of Int. J. Embedded and Real-Time Communication Systems (IJERTCS) and Senior Member of the IEEE. His current research interests include the design and methodological aspects of reliable and secure systems.

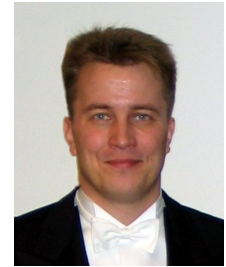

**Jouni Isoaho** received his M.Sc. (Tech) in electrical engineering, and his Lic.Tech. and Dr.Tech. in signal processing at Tampere University of Technology, Finland, in 1989, 1992 and 1995, respectively. During 1994-1998 he worked as an associate professor at Tampere University of Technology, Åbo Akademi University and Royal Institute of Technology (KTH). In 1999 he was invited to the communication systems professorship at the University of Turku, Finland, in 1999. There he heads the communication systems laboratory. His research interests include future communication system concepts, applications and implementation techniques. His current focus is on self-aware systems, dynamic system implementations, system modelling and analysis techniques, and information security and dependability aspects.

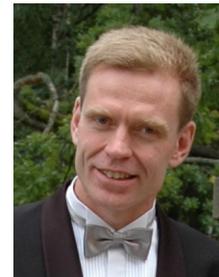